\documentclass[12pt]{iopart}
\usepackage{latexsym}
\usepackage{graphicx}
\begin{document}

\title{
Criticality of the
excess energy cost
due to the unit-flux-quantum external field
for the $(2+1)$D superfluid-insulator transition
}

\author{Yoshihiro Nishiyama}

\address{Department of Physics, Faculty of Science,
Okayama University, Okayama 700-8530, Japan}
%\ead{submissions@iop.org}
\vspace{10pt}
%\begin{indented}
%\item[]August 2017
%\end{indented}

\begin{abstract}

The two-dimensional ($2$D) spin-$S=1$ $XY$ model
was investigated numerically
as
a realization of the 
$(2+1)$D superfluid-Mott-insulator 
(SF-MI)
transition.
The interaction parameters are extended so as to
suppress corrections to finite-size scaling.
Thereby, the external field of a unit flux quantum
($\Phi=2\pi$)
is applied to the 2D cluster
by incorporating the phase factor 
$e^{i\phi_{ij}}$ 
($\phi_{ij}$: gauge angle 
between the $i$ and $j$ sites)
into the hopping amplitudes.
Taking the advantage in that the exact-diagonalization method
allows us
to treat such a complex-valued matrix element,
we evaluated the excess energy cost 
$\Delta E(2\pi)$
due to the magnetic flux $\Phi=2\pi$ explicitly in the 
SF ($XY$) phase.
As a result,
we found that the amplitude ratio
$\rho_s / \Delta E(2\pi)$ 
($\rho_s$: spin stiffness)
makes sense in proximity to the critical point, exhibiting a notable plateau
in the SF-phase side.
The plateau height is estimated, and compared to the related studies.

\end{abstract}

%\maketitle

\section{\label{section1}Introduction}

In two spatial dimensions ($2$D),
the dynamical conductivity $\sigma(\omega)$ becomes a dimensionless
(scale invariant) quantity \cite{Fisher89,Swanson14},
and its Drude weight such as the spin stiffness (helicity modulus)
$\rho_s$
has the same scaling dimension as that of the elementary-excitation masses,
{\it e.g.},
Mott-insulator ($\Delta$)
and Higgs ($m_H$) gaps.
Hence, the critical amplitude ratio
between these quantities
should exhibit a universal behavior
around the superfluid-Mott-insulator phase transition.
Actually, as for the $(2+1)$D boson system,
a variety of critical amplitude ratios
such as $\rho_s/\Delta=0.414$ 
\cite{Rose17}
and $m_H/\Delta=2.2$ 
\cite{Rose15}
were 
calculated via the
renormalization-group 
\cite{Rose17,Rose15,Rancon13,Rancon14,Katan15}
and numerical 
\cite{Hasenbusch08,Gazit13a,Gazit13b,Gazit14,Chen13,Nishiyama15,Nishiyama17}
methods;
see Sec. 4.1 of Ref. \cite{Dupuis21} for a brief overview.
Meanwhile, as to the $(2+1)$D O$(2)$ scalar field theory,
which is relevant to the
superfluid-Mott-insulator phase transition,
the winding-angle-$2\pi$-kink energy
$\Delta E(2\pi)$, namely, vortex's energy,
has been investigated under the 
outward-pointed 
\cite{Delfino19}
and C-periodic 
\cite{Hornung21}
boundary conditions
with the Monte Carlo method.
The former indicates that
the critical amplitude ratio
$\rho_s/\Delta E(2\pi)$ is indeed a universal constant in proximity to the critical point,
whereas the latter revealed
an infrared anomaly due to kink's quantum undulations,
claiming that
the choice of the boundary condition exercises a subtle influence
on
kink's stability.
In fairness,
it has to be mentioned that
in 3D, the character of the kink is arousing much attention
away from the critical point
\cite{Duan94,Baym83,Kopnin91,Thouless07}.
A comprehensive overview will be found in Ref. \cite{Simula18},
where
the {\em kinetic} energy cost of a vortex penetrating 
a finite-thickness plate is considered;
in our simulation, 
the thickness is irrelevant,
because the $(2+1)$D-$XY$ criticality is concerned.

In the present paper,
as a realization of the $(2+1)$D superfluid-Mott-insulator transition,
we consider the 2D spin-$S=1$ $XY$ model.
%%%%%%%%%%%
%%%%%%%%%%%
%%%%%%%%%%%
Here,
the external field of a unit flux quantum 
($\Phi=2\pi$) is applied to the rectangular cluster uniformly
by incorporating the phase factor
$e^{i\phi_{ij}}$
($\phi_{ij}$: gauge angle between the $i$ and $j$ sites)
into the hopping amplitudes.
Taking the advantage in that the exact-diagonalization method
allows us to treat such a complex-valued matrix element,
we evaluated the excess energy cost $\Delta E(2\pi)$
due to $\Phi=2\pi$ explicitly (without performing the inverse Laplace transformation).
A key ingredient of our approach is that 
the
finite-size-scaling behavior is improved 
by extending and adjusting the interaction parameters.
Thereby, with the aid of the finite-size-scaling analysis,
we show that the excess energy cost $\Delta E(2\pi)$
obeys the 3D-$XY$ universality class,
and the critical amplitude ratio $\rho_s/\Delta E(2\pi)$
takes a constant value
in the $XY$ phase.

\begin{figure}
\includegraphics[width=7cm]{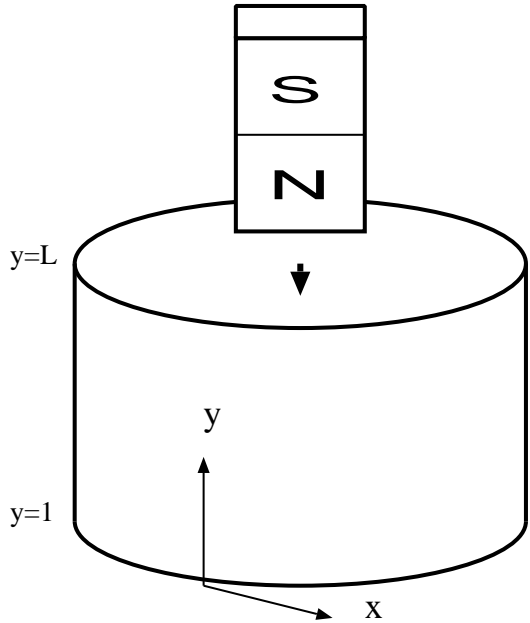}%
\caption{\label{figure1}
We consider the quantum $XY$ model (\ref{Hamiltonian})
on the rectangular cluster as a realization of the
$(2+1)$-dimensional superfluid-Mott-insulator transition
	\cite{Roscilde07}.
We impose the periodic (open) boundary condition
as to the $x$ ($y$) direction,
and hence,
the rectangular cluster forms a cylindrical surface.
Inserting the bar magnet into the cylinder,
we apply 
the magnetic flux $\Phi=2\pi$ per rectangular cluster 
perpendicular to the cylindrical surface
a la Landau gauge (\ref{vector_potential}).
}
\end{figure}

As mentioned above,
we consider the spin-$S=1$ $XY$ model instead of treating
the soft-core boson model directly;
namely,
boson's
creation and annihilation operators are regarded as quantum-spin's ladder operators
\cite{Matsubara56,Roscilde07}.
To be specific, 
the Hamiltonian for the $S=1$ $XY$ model is given by
\begin{eqnarray}
{\cal H} &=&
-\frac{J_{NN}}{2}
\sum_{\langle ij \rangle}
(e^{i\phi_{ij}}S^+_iS^-_j+e^{-i\phi_{ij}}S^-_iS^+_j)
-\frac{J_{NNN}}{2}
\sum_{\langle \langle ij \rangle \rangle}
(e^{i\phi_{ij}}S^+_iS^-_j+e^{-i\phi_{ij}}S^-_iS^+_j)  \nonumber \\
& &
\label{Hamiltonian}
+D \sum_{i=1}^{N}(S^z_i)^2
	+D_{\Box} \sum_{ [ijkl]  } (S^z_i+S^z_j+S^z_k+S^z_l)^2
.
\end{eqnarray}
Here,
the quantum $S=1$ spin
${\bf S}_i$
is placed at each square-lattice point 
$i=1,2,\dots,N$.
The position vector ${\bf r}_i$ of each site $i$
is given by the 2D Cartesian coordinates
${\bf r}_i=(x,y)$ with
$x,y=1,2,\dots,L$ ($N=L^2$).
The periodic (open) boundary condition is imposed along
the $x$ ($y$) direction.
Hence, the $L\times L$ cluster
forms the cylindrical surface, as shown in
Fig.
\ref{figure1}.
In Eq. (\ref{Hamiltonian}),
the summations,
$\sum_{\langle ij \rangle}$,
$\sum_{\langle\langle ij\rangle\rangle}$, and
$\sum_{[ijkl]}$,
run over all possible
nearest-neighbor,
$\langle ij\rangle$,
next-nearest-neighbor,
$\langle\langle ij\rangle\rangle$, 
and
plaquette,
$[ijkl]$, spins, respectively.
The parameters,
$J_{NN}$, $J_{NNN}$, and $D_\Box$, denote
the respective coupling constants.
The gauge twist angle $\phi_{ij}$ is mentioned afterward.
The remaining parameter $D$ stands for the single-ion anisotropy.
Therefore, in the language of boson,
the first two terms of the Hamiltonian
(\ref{Hamiltonian}) correspond to boson's kinetic energy,
whereas the $D$ and $D_\Box$ terms are the repulsive interactions
among the on-site and intra-plaquette bosons, respectively. 
Therefore, the former (latter) enhances the
superfluid (Mott insulator) phase.

%figure1

The magnetic flux is applied by inserting a bar magnet into the cylinder;
see Fig. \ref{figure1}.
The gauge angle $\phi_{ij}$
is set to 
\begin{equation}
\label{gauge_angle}
\phi_{ij}
%={\bf A} \left(\frac{{\bf r}_i+{\bf r}_j}{2}\right) \cdot ({\bf r}_j-{\bf r}_i)
=\int_{{\bf r}_i}^{{\bf r}_j} {\bf A}({\bf r})\cdot d{\bf r}
.
\end{equation}
Here, 
the vector potential 
${\bf A}(x,y)$ 
is given by the expression
\begin{equation}
\label{vector_potential}
	{\bf A}(x,y)=
\left(\frac{ y-\frac{1+L}{2} }{L(L-1)}\Phi
    ,0
\right)
,
\end{equation}
(Landau gauge)
with the flux $\Phi$
threatening the rectangular cluster as a whole.
Hence, the unit-flux-quantum external field is realized by the
setting
$\Phi=2\pi$.

As mentioned above,
the interaction parameters
$(J_{NN},J_{NNN},D,D_\Box)$ are optimized
in order to improve the finite-size-scaling behavior.
Namely,
we survey the subspace
\begin{equation}
\label{interaction_parameter_subspace}
 (J_{NN},J_{NNN},D,D_\Box)
=(jJ_{NN}^*,jJ_{NNN}^*,(2-j)D^*,D_\Box^*)
,
\end{equation}
parameterized by the variable $j$.
Here, the optimal critical point 
\cite{Nishiyama08}
\begin{equation}
\label{optimal_critical_point}
(J_{NN}^*,J_{NNN}^*,D^*,D_\Box^*)=(
0.1582 %42810160
,
0.05856 % 1393564
,
0.957     %  (25)
,
0.1003 % 5104389)
	)
\end{equation}
was determined with the ordinary finite-size-scaling method
combined with the
real-space decimation so as to get rid of the irrelevant interaction 
terms \cite{Hasenfratz98},
and attain suppressed corrections to scaling.
The critical point $j_c$ is thus given by
\begin{equation}
\label{critical_point}
j_c=1
,
\end{equation}
at which the set of parameters, $(J_{NN},J_{NNN},D,D_\Box)$,
reduces to
that of the critical point (\ref{optimal_critical_point}).

A schematic drawing of the ground-state phase diagram 
is shown in Fig. \ref{figure2}.
For large $j>1(=j_c)$, the $XY$-ordered phase is realized,
whereas in $j<1$, the paramagnetic phase extends \cite{Wang05}.
The criticality at $j_c=1$ (\ref{critical_point})
belongs to the 3D-$XY$ universality class
\cite{Roscilde07}.
In the language of boson, the $XY$ (paramagnetic) phase
corresponds to the superfluid (Mott insulator) phase \cite{Roscilde07}.
As mentioned above, the gauge flux $\Phi$
(\ref{vector_potential}) and the $XY$ order
conflict each other, and hence, the excess energy cost 
should take a non-zero
value in the $XY$ phase.

%figure2

The rest of this paper is organized as follows.
In Sec. \ref{section2}, the simulation results for the $XY$ model 
(\ref{Hamiltonian}) are presented.
Details of the finite-size scaling are explained as well.
In Sec. \ref{section3},
we present the the summary and discussions.

\section{\label{section2}Numerical results}

In this section we present the numerical results
for the two-dimensional $XY$ model (\ref{Hamiltonian})
subjected to the gauge flux $\Phi$ (\ref{vector_potential}).
We employed the exact-diagonalization method,
which enables us to treat
the
gauge-twisted complex-valued matrix element,
and evaluate the excess energy cost due to $\Phi=2\pi$
explicitly.
In our preliminary survey, 
we found that irrespective of the value of $\Phi$,
the ground state belongs to the 
$(S^z_{tot},k_x)=(0,0)$ sector
with the total longitudinal spin moment, $S^z_{tot}$, and
$x$ component of the wave vector, $k_x$.
Within this subspace,
the numerical diagonalization was performed.
Hence, the translational motion of the $\Phi=2\pi$ kink 
is prohibited {\it a priori}, even though 
the translational drift costs very little energy
with a quadratic (soft mode) dispersion relation, $\propto k_x^2$;
it is an advantage of the exact-diagonalization method
in that the quantum number $k_x$ of kink's drift
is under control.

\subsection{\label{section2_1}Finite-size scaling of the excess energy cost $\Delta E(2\pi)$}

In this section we consider the excess energy cost
\begin{equation}
\label{excess_energy_cost}
\Delta E (2\pi)= E_0(2\pi)-E_0(0)
,
\end{equation}
with the ground state energy $E_0(\Phi)$ of the Hamiltonian ${\cal H}$
(\ref{Hamiltonian})
under the gauge flux $\Phi$ (\ref{vector_potential}).
The gauge flux should create a winding-angle-$\Phi$ kink within the $XY$ order.

\begin{figure}
\includegraphics[width=13cm]{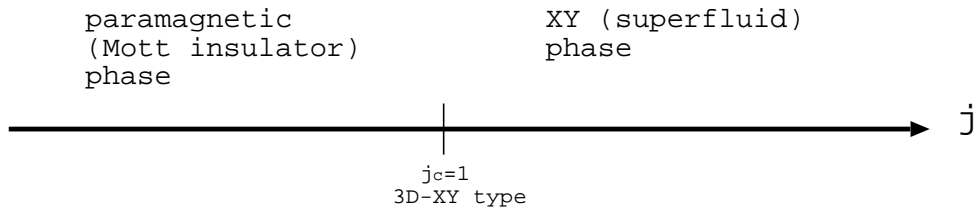}%
\caption{\label{figure2}
A schematic ground-state phase diagram for the two-dimensional $XY$ model
(\ref{Hamiltonian})
with the coupling constants parameterized by $j$ 
(\ref{interaction_parameter_subspace})
is presented.
%here, the
%the  gauge flux $\Phi$ is turned off, $\Phi=0$.
For large (small) $j$, the $XY$ (paramagnetic)
phase is realized.
	In the boson language \cite{Roscilde07},
each phase corresponds
to the superfluid (Mott insulator) phase.
The critical point at $j_c=1$ (\ref{critical_point})
belongs to 
the
$3$D-$XY$ universality class \cite{Wang05}.
The scaling behavior is improved 
\cite{Hasenfratz98}
by
extending and adjusting the 
interaction parameters
as in Eq. 
(\ref{interaction_parameter_subspace}) \cite{Nishiyama08}.}
\end{figure}

\begin{figure}
\includegraphics[width=13cm]{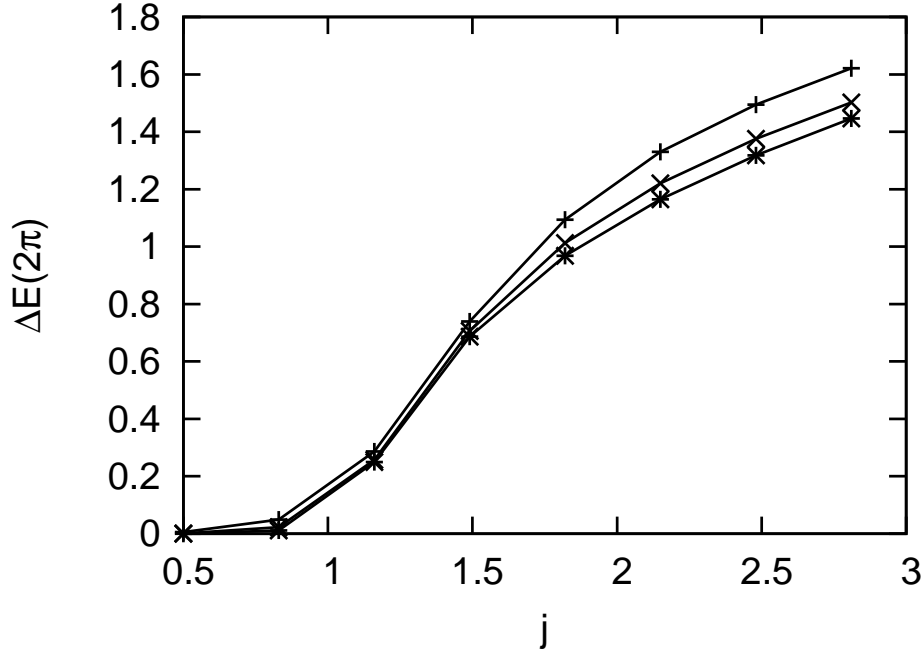}%
\caption{\label{figure3}
The excess energy cost 
$\Delta E(2\pi)$ (\ref{excess_energy_cost})
due to the gauge flux $\Phi=2\pi$
is presented for the interaction parameter $j$ 
(\ref{interaction_parameter_subspace})
and the system sizes,
($+$) $L=3$,
($\times$) $4$, and
($*$) $5$.
In the $XY$ (superfluid) phase $j>j_c(=1)$, the excess energy cost
	$\Delta E (2\pi)$ develops.
%Such a feature is interpreted as the
%Meissner effect
%in the boson language.
}
\end{figure}

In Fig. \ref{figure3},
we present the excess energy cost $\Delta E(2\pi)$
(\ref{excess_energy_cost})
for various values of the interaction parameter $j$
(\ref{interaction_parameter_subspace})
and the system sizes,
($+$) $L=3$
($\times$) $4$, and
($*$) $5$.
The excess energy cost
appears to
develop in the $XY$ phase, $j>j_c(=1)$,
whereas it vanishes in the paramagnetic phase, $ j <j_c$.
As anticipated, the excess energy cost reflects
an elasticity of the $XY$ order.
Actually, 
this energy cost comes from the conflict between 
the superfluid state and the gauge flux
reminiscent of the Meissner effect;
the correspondence between the spin and boson pictures
is shown
in the chart, 
Fig. \ref{figure2}.
Strictly speaking, unlike the Meissner effect,
the flux takes a constant value irrespective of $L$,
introducing a kink within the system.
The texture of the kink is not pursued here,
because such a snapshot picture is not available in the 
exact-diagonalization scheme.

%figure3

\begin{figure}
\includegraphics[width=13cm]{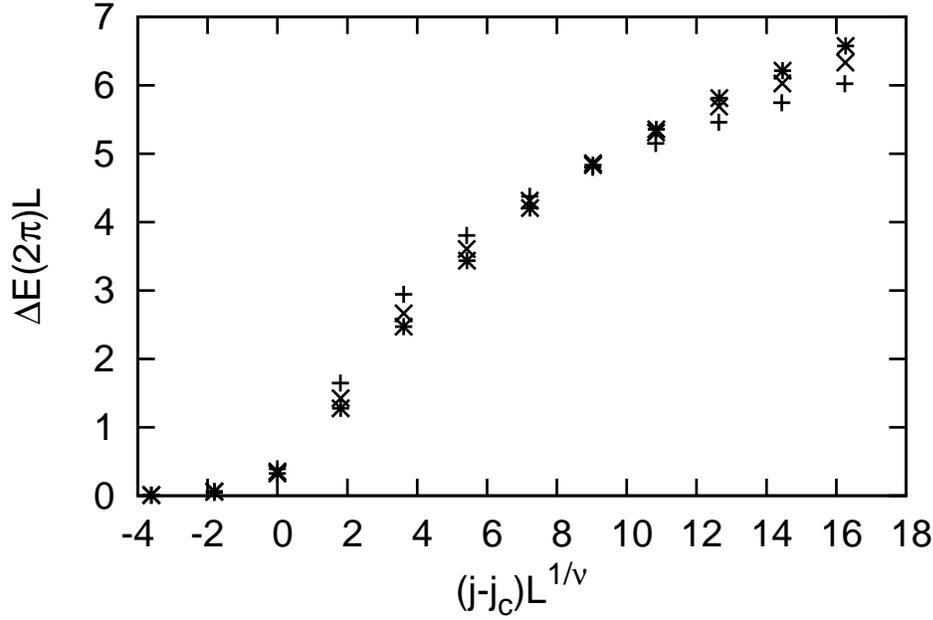}%
\caption{\label{figure4}
The scaling plot, $(j-j_c)L^{1/\nu}$-$\Delta E(2\pi)L$, of $\Delta E(2\pi)$
is shown for various system sizes,
($+$) $L=3$,
($\times$) $4$, and
($*$) $5$.
Here, the scaling parameters,
namely,
the critical point and correlation-length critical exponent,
are set to
$j_c=1$ (\ref{critical_point}) and
$\nu=0.6717$ (3D-$XY$ universality class) \cite{Campostrini06,Burovski06}, respectively.
The simulation data collapse into the scaling curve
satisfactorily,
indicating that the simulation results already 
enter into the scaling regime.
The scaling behavior appears to be improved
\cite{Hasenfratz98}
by the extention of the interaction parameters
as in Eq. (\ref{interaction_parameter_subspace}).
}
\end{figure}

We turn to the analysis of
the criticality of
the excess energy cost $\Delta E(2\pi)$.
In Fig. \ref{figure4},
we present the scaling plot,
$(j-j_c)L^{1/\nu}$-$\Delta E(2\pi) L$,
of $\Delta E(2\pi)$
for various system sizes,
($+$) $L=3$
($\times$) $4$, and
($*$) $5$.
The underlying idea behind the ordinate-axis scale,
$\Delta E(2\pi)L$, is as follows.
We made an assumption that $\Delta E(2\pi)$ should have the
same scaling dimension as that of the mass gap $m$.
Because
the mass gap scales as the inverse correlation length $m \sim \xi^{-1}$ 
(along the imaginary-time direction),
the expression $\Delta E (2\pi) L$ should
be a scale-invariant quantity owing to the scaling hypothesis,
$\xi \sim L$.
On the one hand,
the scale invariance of the abscissa scale, $(j-j_c)L^{1/\nu}$, 
follows immediately from 
the definition of the correlation-length critical exponent
$\nu$, {\it i.e.},
$\xi\sim|j-j_c|^{-\nu}$.
The scaling parameters, $j_c$ and $\nu$, are set to
$j_c=1$ (\ref{critical_point}),
and $\nu=0.6717$ \cite{Campostrini06,Burovski06},
respectively.
The latter is taken from the value of the 3D-$XY$ universality class,
as mentioned in Introduction.
We stress that
there is no {\it ad hoc} adjustable parameter
undertaken
in the present scaling analysis.

%figure4

In Fig. \ref{figure4}, we see that the scaled $\Delta E(2\pi)$
data fall into the scaling curve 
satisfactorily,
indicating that the quantity $\Delta E(2\pi)$ 
obeys
the 3D-$XY$ universality class.
In other words, 
the simulation data for $\Delta E(2\pi)$ already enter into the scaling regime.
Actually,
owing to the fine tuning of the interaction parameters 
as in Eq. (\ref{interaction_parameter_subspace}),
corrections to scaling are eliminated \cite{Hasenfratz98}
to a considerable extent.
Encouraged by this finding, we further explore the criticality of
the
spin stiffness in the next section.   %%, which is a natural counterpart of $\Delta E(2\pi)$

Last, we address a number of remarks.
First,
the scaling plot, Fig. \ref{figure4}, indicates that the 
winding-angle-$\Phi=2\pi$ gauge field
(\ref{vector_potential})
indeed creates a
point-like excitation,
because
the excess energy cost
$\Delta E(2\pi)$
has the same scaling dimension as the excitation mass, as confirmed above.
In fact,
away from the critical point $j \approx 1.5(>j_c)$,
in Fig. \ref{figure3},
kink's energy $\Delta E(2\pi)$ appears to be almost $L$-independent.
Hence, kink's size should be sufficiently smaller than $L$,
at least, away from $j_c$.
Such a feature supports
the mean-field (Bogoliubov-de Gennes) analysis
\cite{Simula18}, which 
states that kink's core is responsible for the kinetic mass.
Last, in $(2+1)$ dimensions,
the choice of the boundary condition 
is ``problematical'' 
\cite{Hornung21} in regard to the vortex stability.
In our setting,
as depicted in Fig. \ref{figure2},
the system is translation invariant along the $x$ axis,
whereas the open boundary condition is imposed as to the $y$ direction.
In this sense, our setting is reminiscent of that of Ref. \cite{Delfino19},
where the outward-pointed boundary condition is imposed for 
all edges of the finite-size cluster, and vortex's mass is appreciated properly.
Namely, the open boundary condition contributes to the stabilization of
the vortex \cite{Hornung21}.
Moreover,
the numerical diagonalization was performed within the zero-momentum
($k_x=0$) subspace,
and thus,
the drift
along the $x$ direction 
is prohibited
{\it a priori}.
This treatment may also suppress the drift of the kink.

\subsection{\label{section2_2}Finite-size scaling of the spin stiffness $\rho_s$}

In this section we present the result for the spin stiffness $\rho_s$.
For that purpose, in this section,
the vector potential is set to the spatially uniform one
\begin{equation}
\label{stiffness_vector_potential}
{\bf A}({\bf r})=
\left(\frac{\theta}{L},0 \right)
,
\end{equation}
with the gauge twist angle $\theta$ through the boundary condition along the
$x$ direction.
This situation is realized by a sufficiently long bar magnet
with the flux $\theta$ threatening through the cylinder.
Clearly,
this geometrical arrangement resembles that of Fig. \ref{figure1}.
Accordingly,
the gauge twist angle $\phi_{ij}$ in the Hamiltonian (\ref{Hamiltonian})
has to be set in the same way as in Eq. (\ref{gauge_angle}).
We are now able to impose
the periodic boundary condition for both $x$ and $y$ directions,
because the vector potential (\ref{stiffness_vector_potential})
is a constant one.
Thereby, the spin stiffness was calculated as the elastic
constant with respect to the 
distortion $\theta$
as
\begin{equation}
	\label{spin_stiffness}
\rho_s = 
\left.
\frac{\partial^2 E_0}{\partial \theta^2}
\right|_{\theta=0}
,
\end{equation}
with the ground state energy $E_0$  of the Hamiltonian (\ref{Hamiltonian})
under the gauge (\ref{stiffness_vector_potential}).
In the $XY$ phase, the spin stiffness should take
a non-zero value because of the elasticity of the $XY$ order.

\begin{figure}
\includegraphics[width=13cm]{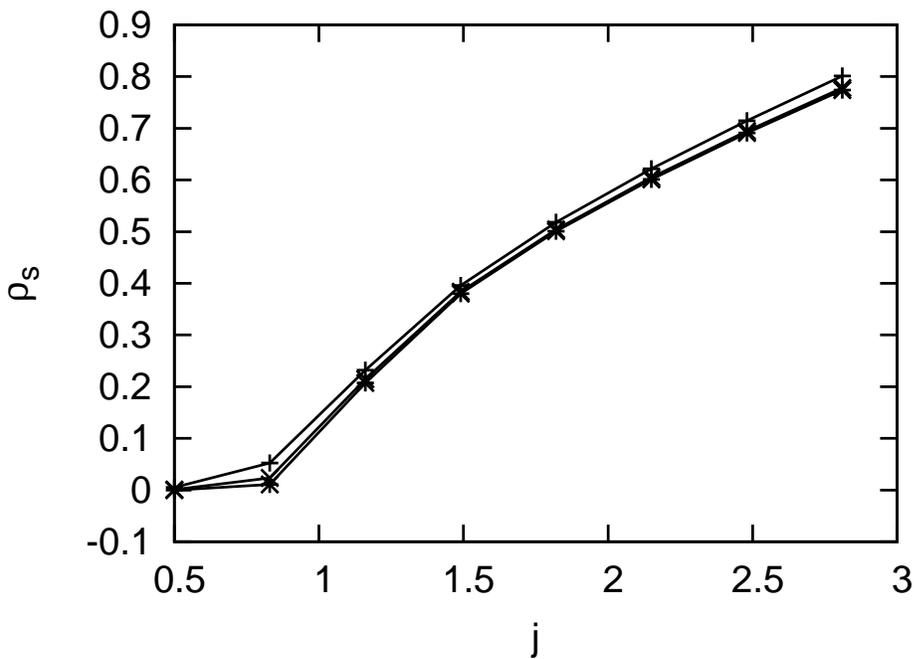}%
\caption{\label{figure5}
The spin stiffness 
$\rho_s$ (\ref{spin_stiffness})
is presented for the interaction parameter $j$ 
(\ref{interaction_parameter_subspace})
and various system sizes,
($+$) $L=3$,
($\times$) $4$, and
($*$) $5$.
In the $XY$ (superfluid) phase $j>j_c(=1)$, the
spin stiffness develops
in a way reminiscent of
the excess energy
cost $\Delta E(2\pi)$, as presented in Fig. \ref{figure3}.
}
\end{figure}

To begin with,
in Fig. \ref{figure5},
we present the spin stiffness $\rho_s$ (\ref{spin_stiffness})
for various values of the interaction parameter
$j$ and 
system sizes,
($+$) $L=3$.
($\times$) $4$, and
($*$) $5$.
The spin stiffness develops in the $XY$ phase, $j>j_c(=1)$,
whereas it is suppressed in the paramagnetic phase, $j<j_c$.
Such a character resembles that of the excess energy cost
$\Delta E(2\pi)$, 
and actually,
the behaviors of Fig. \ref{figure3} and \ref{figure5} look alike.
Hence, the spin stiffness should be a good counterpart of the excess
energy cost, and the critical amplitude ratio 
between these quantities is investigated in the next section.

%figure5

\begin{figure}
\includegraphics[width=13cm]{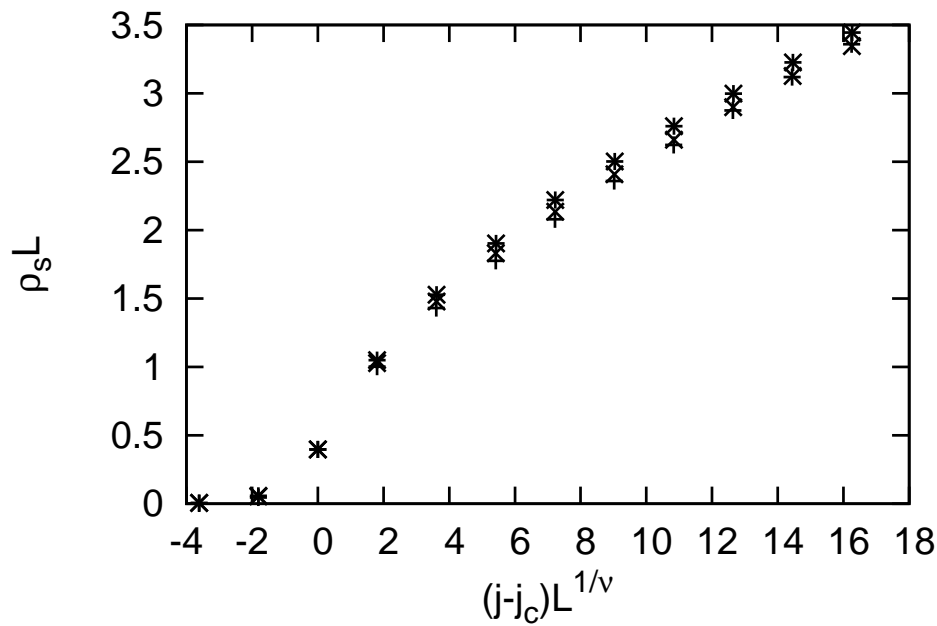}%
\caption{\label{figure6}
The scaling plot, $(j-j_c)L^{1/\nu}$-$\rho_s L$, of $\rho_s$
is shown for various system sizes,
($+$) $L=3$,
($\times$) $4$, and
($*$) $5$.
Here, the scaling parameters,
namely,
the critical point $j_c$
and the correlation-length critical exponent $\nu$,
are the same as those of Fig. \ref{figure4}.
Even without any {\it ad hoc} adjustable parameters,
the scaling data for $\rho_s$ fall into the scaling curve
satisfactorily,
indicating that the criticality indeed belongs to
the 3D-$XY$ universality class.
}
\end{figure}

Aiming to examine the critical behavior of the spin stiffness,
we present the scaling plot,
$(j-j_c)L^{1/\nu}$-$\rho_s L$, 
% of $\rho_s$
for various system sizes,
($+$) $L=3$.
($\times$) $4$, and
($*$) $5$,
in Fig. \ref{figure6}.
The underlying idea behind the scaling plot is as follows.
The ordinate axis scale $\rho_s L$
is invariant in two spatial dimensions
according to the scaling argument \cite{Fisher89};
therefore, the spin stiffness has the same scaling dimension as that
of the elementary excitation gap as well as $\Delta E(2\pi)$.
On the one hand,
the abscissa scale $(j-j_c)L^{1/\nu}$ is the same as that of Fig. \ref{figure4}.
Additionally,
the scaling parameters, $j_c$ and $\nu$, are 
the same as those of Fig. \ref{figure4}.

%figure6

The scaling data in Fig. \ref{figure6} overlap each other
satisfactorily, confirming that the criticality is
under the reign of the 3D-$XY$ universality class.
We stress that
the scaling parameters,
$j_c$ and $\nu$, are identical to those of Fig. \ref{figure4}.
and there is no adjustable scaling parameters.
Such a feature indicates that 
the optimized interaction parameters
(\ref{interaction_parameter_subspace})
indeed contribute to the suppression of 
corrections to scaling
\cite{Hasenfratz98}
even for $\rho_s$.

This is a good position to address a number of remarks.
First, as demonstrated above,
the spin stiffness (\ref{spin_stiffness})
is less computationally demanding
in the exact-diagonalization approach.
With the world-line Monte Carlo method,
the spin stiffness is evaluated systematically by the
winding number of the world lines across the boundary condition
\cite{Pollock87};
this idea, however, 
does not apply to the Monte Carlo method
based on the O$(2)$-scalar-field representation.
Last,
the spin stiffness $\rho_s$ is 
not a mere theoretical concept,
because it is observable
experimentally
\cite{Corson99,Crane07,Sherson10}.
Therefore, 
via $\rho_s$,
other quantities can also be appreciated
indirectly
by relying on 
the critical amplitude ratios
\cite{Rose17,Rose15,Rancon13,Rancon14,Katan15,
Hasenbusch08,Gazit13a,Gazit13b,Gazit14,Chen13,Nishiyama15,Nishiyama17}.
In the next section, we follow this idea, choosing the excess energy cost
$\Delta E(2\pi)$ as the denominator of critical amplitude ratio.

\subsection{\label{section2_3}Critical amplitude ratio 
$\rho_s / \Delta E(2\pi)$}

In this section, we turn to the analysis of
the amplitude ratio $\rho_s / \Delta E(2\pi)$,
following
the preliminaries in Sec. \ref{section2_1} and \ref{section2_2}.

\begin{figure}
\includegraphics[width=13cm]{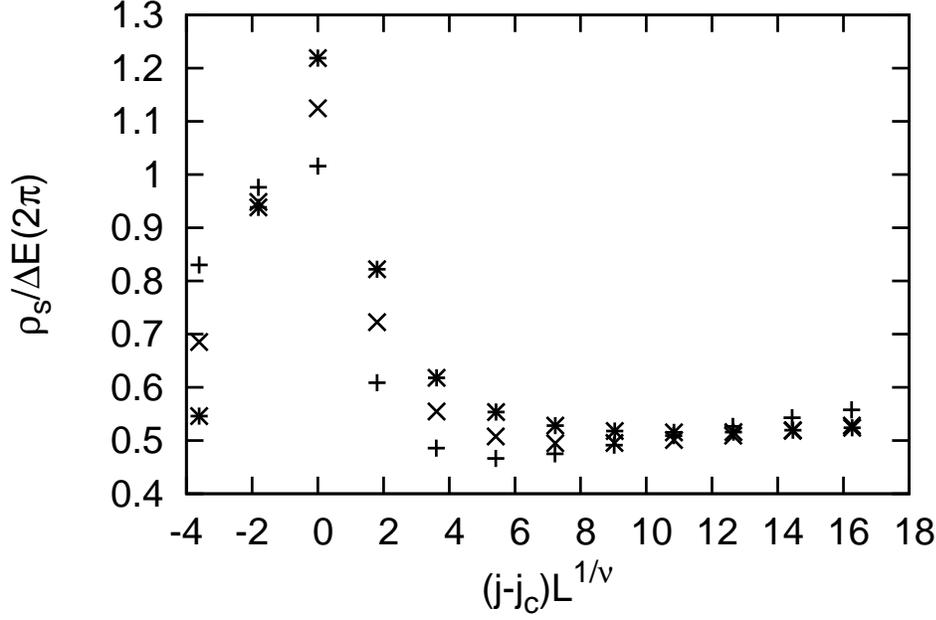}%
\caption{\label{figure7}
The scaling plot, $(j-j_c)L^{1/\nu}$-$\rho_s / \Delta E(2\pi)$, of 
the amplitude ratio % $\Delta E(2\pi)/\rho_s$
is shown for various system sizes,
($+$) $L=3$,
($\times$) $4$, and
($*$) $5$.
Here, the scaling parameters,
namely,
the critical point $j_c$
and the correlation-length critical exponent $\nu$,
are the same as those of Fig. \ref{figure4}.
The amplitude ratio exhibits a notable plateau in the $XY$-phase 
side, $(j-j_c)L^{1 / \nu}\approx 10 $,
indicating that the amplitude ratio takes a universal constant in
proximity to the critical point.
}
\end{figure}

In Fig. \ref{figure7},
we present the scaling plot,
$(j-j_c)L^{1/\nu}$-$\rho_s / \Delta E(2\pi)$,
of the amplitude ratio
for various system sizes,
($+$) $L=3$
($\times$) $4$, and
($*$) $5$.
Here,
the scaling parameters, 
$j_c$ and $\nu$,
are the same as those of Fig. \ref{figure4}.
The ordinate axis $\rho_s /\Delta E(2\pi)$ is scaling invariant.
Actually, in Sec. \ref{section2_1} and \ref{section2_2},
it was found that
the numerator and denominator, $\rho_s$ and $\Delta E(2\pi)$,
respectively, have the same scaling dimensionality,
$L^{-1}$. The abscissa scale is the same as that of Fig. \ref{figure4}.

%figure7

In Fig. \ref{figure7},
the scaled data overlap each other to form a plateau
in the $XY$ phase, $(j-j_c)L^{1/\nu} > 0 $.
Such a feature
supports that the amplitude ratio $\rho_s/\Delta E(2\pi)$
indeed takes a universal constant in this domain.
The plateau height is roughly estimated as 
$\rho_s / \Delta E(2\pi)\approx 0.5$
around $(j-j_c)L^{1/\nu}\approx 10 $.
Hence, for sufficiently large $L\to\infty$, this plateau regime $j$ approaches
toward the critical point as $j-j_c \to 0^+$.
On the one hand, 
in the paramagnetic phase, $(j-j_c)L^{1/\nu} < 0$,
a rapid convergence to $\rho_s/\Delta E(2\pi) \to 0 $ is observed,
while in close vicinity of the critical point,
$(j-j_c)L^{1/\nu} \approx 0$,
a steep development of a  peak is seen.
It would be reasonable
that the ratio $\rho_x /\Delta E(2\pi)$ exhibits
such singular behaviors
in the paramagnetic phase, where both numerator and denominator
go to
$ \rho_s,\Delta E(2\pi) \to 0$ simultaneously, as $L \to \infty$.

\begin{figure}
\includegraphics[width=13cm]{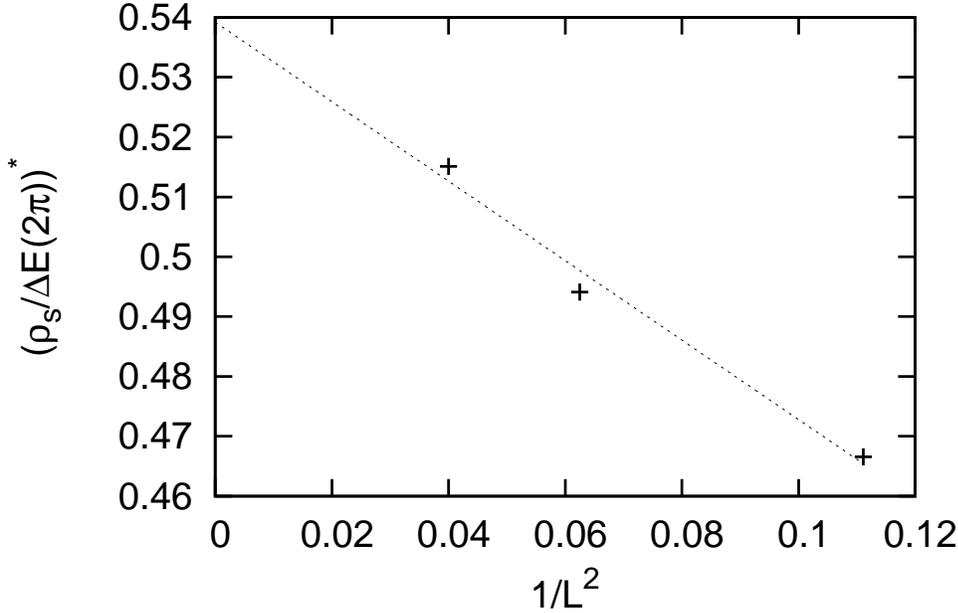}%
\caption{\label{figure8}
The approximate amplitude ratio 
	$(\rho_s / \Delta E(2\pi))^*$ (\ref{approximate_amplitude_ratio})
is plotted for $1/L^2$.
The least-squares fit to these data yields 
an estimate $\rho_s / \Delta E(2\pi)=0.539(7)$
in the thermodynamic limit, $L\to\infty$.
A possible systematic error is considered in the text.
}
\end{figure}

In order to estimate the plateau height, namely,
the amplitude ratio, precisely,
in Fig. \ref{figure8},
we present the approximate amplitude ratio
\begin{equation}
\label{approximate_amplitude_ratio}
(  \rho_s / \Delta E(2\pi))^*(L)=
\left.
	\frac{\rho_s} {\Delta E(2\pi)}
\right|_{j=j_c^*(L)},
\end{equation}
for $1/L^2$.
Here, the approximate critical point $j^*_c(L)$
denotes 
the extremal point
\begin{equation}
\left.
	\partial_j \frac{\rho_s}{\Delta E(2\pi)}
\right|_{j=j^*(L)}
=0 ,
\end{equation}
of the above-mentioned plateau
for each $L$.
The least-squares fit to the data
in Fig. \ref{figure8}
yields an estimate 
$\rho_s / \Delta E(2\pi)= 0.539(7)$ in the thermodynamic limit, $L\to\infty$.
In order to applicate a possible systematic error,
replacing the abscissa scale of Fig. \ref{figure8} with $1/L$,
we carried out an alternative extrapolation analysis.
Thereby, we arrived at a result,
$\rho_s / \Delta E(2\pi)=0.585(6)$.
The deviation $\approx 0.05$
from the aforementioned estimate $0.539$ seems to dominate the least-squares-fitting error,
$\approx 0.007$.
Hence, 
regarding the former as the dominant source of uncertainty,
we estimate the amplitude ratio
as 
\begin{equation}
\label{amplitude_ratio}
\rho_s / \Delta E(2\pi)=0.54(5)
.
\end{equation}

%figure8

\begin{figure}
\includegraphics[width=13cm]{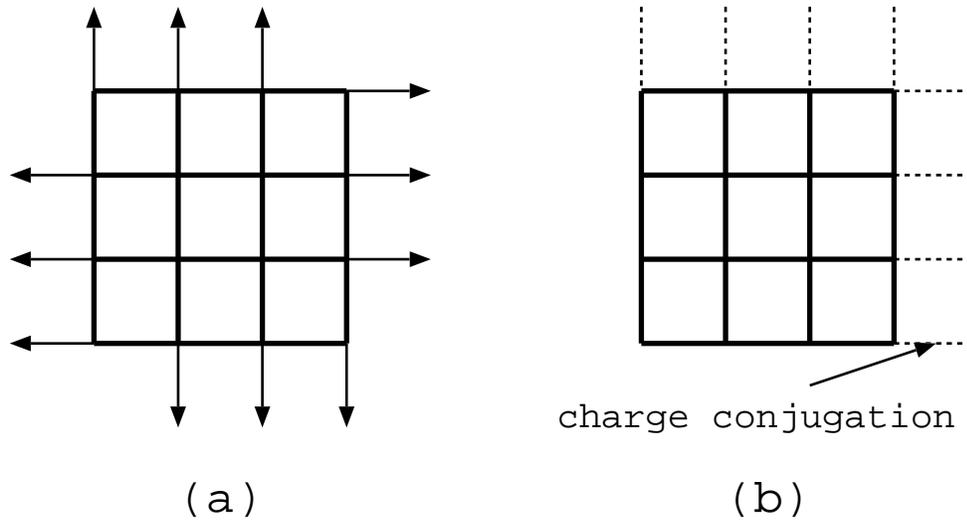}%
\caption{\label{figure9}
Schematic drawings for the (a)
outward-pointed \cite{Delfino19}
	and (b)
	C-periodic \cite{Hornung21}
	boundary conditions are presented.
	In the former, the boundary spins are directed outward,
	whereas in the latter,
	there are imposed
	the periodic boundary conditions with the charge-conjugation twists,
	which ``leave translation invariance intact'' \cite{Hornung21}.
}
\end{figure}

We recollect a number of related studies.
As mentioned in Introduction,
according to the Monte Carlo simulation of
the 3D classical O$(2)$ scalar field theory under the outward-pointed boundary condition \cite{Delfino19},
the amplitude ratio was estimated 
as
\begin{equation}
\label{Delfino}
\rho_s/\Delta E(2\pi) \approx 0.4
.
\end{equation} 
This final result was obtained, relying
on
the preceding Monte Carlo data,
$\rho_s/\Delta=0.411(2)$
($\Delta$: Mott-insulator gap)
\cite{Hasenbusch08}.
This estimate (\ref{Delfino})
appears to
lie out of the error margin of ours (\ref{amplitude_ratio}).
As mentioned above, this result (\ref{Delfino})
was evaluated under the outward-pointed boundary condition \cite{Delfino19};
%%%%
%%%%
%%%%
%%%%
%%%%
see Fig. \ref{figure9} (a).
This boundary condition 
was implemented in such a way that
rectangular-cluster-boundary spins
are enforced to point outward,
and thereby, 
an estimate,
$\rho_s /m_V \approx 0.2 $ ($m_V$: vortex mass)
was obtained.
This mass $m_V$ corresponds
to
$m_V(=\Delta E(4\pi))=2 \Delta E (2\pi)$,
because of
the winding-angle-$(\frac{2\pi}{4} \times 4)$ stress at each corner of the rectangular cluster
and
the winding-angle-$2 \pi$ defect in the midst of the cluster.
Hence, the aforementioned relation,
$\rho_s / m_V \approx 0.2$ \cite{Delfino19},
admits an expression
$\rho_s / \Delta E (2\pi)\approx 0.4$, Eq. (\ref{Delfino}), to adapt our notation.
%%
%%
%%%%
%%%%
%%%%
%%%%
%%%%
%%%%
The discrepancy between the preceding result (\ref{Delfino})
and
ours (\ref{amplitude_ratio})
might be attributed to the details of the boundary conditions undertaken.
Actually, in our setting, Fig. \ref{figure1}, the periodic boundary condition is imposed
on the $x$ direction.
%%%%
%%%%
%%%%
%%%%
%%%%
%%%%
%%%%
%%%%
%%%%
On the contrary,
the Monte Carlo simulation under
the C-periodic boundary condition \cite{Hornung21}
(see Fig. \ref{figure9} (b))
revealed
an infrared anomaly for the vortex mass, 
claiming that
the choice of the boundary condition
leads to significant consequences as to this problem.
%%%%
%%%%
%%%%
%%%%
Actually, according to this study \cite{Hornung21},
the vortex energy shows logarithmic divergences, as $L$ increases.
In our setting, on the contrary,
kink's energy appears to be almost $L$-independent around 
$j\approx 1.5$, as Fig. \ref{figure3} indicates;
%%%%
namely, kink's stress energy concentrates in its core.
%%%%
%%%%
%%%%
%%%%
That is, the periodic boundary condition promotes the infrared undulation of the kink.
In our approach,
the open boundary condition along the $y$ axis
[because of the Landau gauge (\ref{vector_potential})]
and Hilbert-space's restriction within $k_x=0$ 
are responsible for the stabilization of the kink, as argued in
Sec. \ref{section2_1}.

%%%%
%%%%
%%%%
%%%%
We address a number of remarks.
First,
the abscissa scale $1/L^2$ in Fig. \ref{figure8}
comes from the index $\omega_2\approx 2$ \cite{Campostrini06}
for next-to-leading scaling corrections.
Because the leading ones are suppressed
\cite{Nishiyama08,Hasenfratz98}
by adjusting the interaction parameters to Eq. (\ref{optimal_critical_point}),
the universal quantity such as $\rho_s/\Delta E(2\pi)$
should obey the power-law behavior with this index $\omega_2\approx 2$
for the small-$L$ regime at least.
Last, a peculiarity of the amplitude ratio $\rho_s/\Delta E(2\pi)$ is that
in the opposite side (paramagnetic phase), the value is rather scattered;
in contrast, Higgs-mass's amplitude ratio
$m_H/\rho_s$, for instance, exhibits less singular behavior,
and actually, it looks alike for both phases \cite{Chen13}.

\section{\label{section3}Summary and discussions}

% (=0.2*2)=0.2  to kaku

The excess energy cost $\Delta E (\Phi)$ 
(\ref{excess_energy_cost}) due to the external flux $\Phi = 2\pi$
in the superfluid phase
was investigated 
by  means of the exact-diagonalization method,
with which one is able to
treat the gauge-twisted complex-valued matrix elements.
As a realization of the superfluid phase,
we consider 
the quantum $XY$ magnet (\ref{Hamiltonian})
\cite{Roscilde07}
with the extended interactions
(\ref{interaction_parameter_subspace})
to improve the finite-size-scaling behavior \cite{Hasenfratz98},
Actually, the excess energy cost $\Delta E(2\pi)$ 
appears to obey 
the 3D-$XY$ universality class
satisfactorily.
Thereby,
choosing the spin stiffness $\rho_s$ as its counterpart,
we analyzed
the amplitude ratio, $\rho_s / \Delta E(2\pi) $,
postulating that the criticality belongs to the 
3D-$XY$ universality class.
The amplitude ratio $\rho_s / \Delta E(2\pi)$
exhibits a notable plateau around
$(j-j_c)L^{1/\nu} \approx 10$,
and thus,
this plateau regime approaches to the critical point, $j-j_c\to 0^+$, as $L\to\infty$.
The plateau height is estimated as $\rho_s / \Delta E(2\pi) = 0.54(5)$, Eq. (\ref{amplitude_ratio}).
So far, under the 
outward-pointed
\cite{Delfino19}
and 
C-periodic
\cite{Hornung21}
boundary conditions, extensive simulations have been made.
The former estimate 
$\rho_s /\Delta E(2\pi) \approx 0.4$, Eq. (\ref{Delfino}),
appears to lie out of the error margin of ours,
whereas the latter
revealed
an infrared anomaly,
claiming that the choice of the boundary condition is significant.
In the present study,
the open boundary condition along the $y$ axis
and Hilbert-space's restriction within $k_x=0$ 
contribute to the stabilization of $\Delta E(2\pi)$.
Nevertheless,
as demonstrated,
the spin stiffness is of use to 
elucidate
the universal character of $\Delta E(2\pi)$ quantitatively.

Through the duality transformation
\cite{Stone78,Fisher89b,Wen90}, the Mott-insulator state
is interpreted as the condensed state of the vortices,
and the vortex stiffness $\rho_v$ now makes sense.
Noticeably,
this quantity $\rho_v$ is  accessible via the 
Nozi\'eres-Pines formula \cite{Nozieres58}.
It would be tempting to evaluate the ratio $\rho_v / \Delta E(2\pi) $
across the Mott-insulator and superfluid phases as a ``quantitative measure''
\cite{Gazit14}
of deviation from self-duality.
% gazit prl 113 240601 14
This problem is left for the future study.

\ack

This work was supported by a Grant-in-Aid
for Scientific Research (C)
from Japan Society for the Promotion of Science
% from Monbu-Kagakusho, Japan
(Grant No. 20K03767).

%%\References

%\section*{References}

\end{document}